\documentclass{article}
\usepackage{ijcai25}

\usepackage{times}
\usepackage{soul}
\usepackage{url}
\usepackage[hidelinks]{hyperref}
\usepackage[utf8]{inputenc}
\usepackage[small]{caption}
\usepackage{graphicx}
\usepackage{amsmath}
\usepackage{amsthm}
\usepackage{booktabs}
\usepackage{algorithm}
\usepackage{amssymb}
\usepackage{algorithmic}
\usepackage[switch]{lineno}
\usepackage{multirow} 
\title{UniSpeaker: A Unified Approach for Multimodality-driven Speaker Generation}

\author{
Zhengyan Sheng$^1$\and
Zhihao Du$^2$\and
Heng Lu$^2$ \and
Shiliang Zhang$^2$ \and
Zhen-Hua Ling$^{1*}$\\
\affiliations
$^1$University of Science and Technology of China\\
$^2$Speech Lab,  Alibaba Group, China\\
\emails
zysheng@mail.ustc.edu.cn,
\{neo.dzh, sly.zsl\}@alibaba-inc.com,
zhling@ustc.edu.cn
}
\begin{document}

\maketitle

\begin{abstract}
Recent advancements in personalized speech generation have brought synthetic speech increasingly close to the realism of target speakers' recordings, yet multimodal speaker generation remains on the rise.  This paper introduces UniSpeaker, a unified approach for multimodality-driven speaker generation. Specifically, we propose a unified voice aggregator based on KV-Former, applying soft contrastive loss to map diverse voice description modalities into a shared voice space, ensuring that the generated voice aligns more closely with the input descriptions. To evaluate multimodality-driven voice control, we build the first multimodality-based voice control (MVC) 
 benchmark, focusing on voice suitability, voice diversity, and speech quality. UniSpeaker is evaluated across five tasks using the MVC benchmark, and the experimental results demonstrate that UniSpeaker outperforms previous modality-specific models. Speech samples are available at \url{https://UniSpeaker.github.io}.
\end{abstract}

\section{Introduction}
In recent years, the field of speech synthesis has seen remarkable progress \cite{valle,cosyvoice,naturalspeech3}, enabling the generated speech to closely resemble the actual recordings. However, traditional zero-shot speech synthesis still faces limitations in certain scenarios, such as providing voiceovers for virtual characters, where obtaining ideal reference speech is very difficult or even nonexistent \cite{prompttts}.  Therefore, the voice control abilities in generative models need to transition from speaker cloning to speaker generation.
% 因此，生成模型的音色控制能力需要从模仿音色转向创造音色。
Compared to cloning voice based on the reference speech, using other more convenient modalities to express the intentions holds great potential for creating the desired voice characteristics.
% Compared to reference speech, \cite{prompttts2, face2speech,speech2face}. In the absence of reference speech, utilizing other modalities allows for more flexible and convenient control over voice characteristics.
% Hence, multimodal voice description-based speech generation, which involves generating corresponding voice characteristics from natural text descriptions, face images, or other modalities, presents a promising approach.

Recently, several studies \cite{prompttts++,promptspeaker,facevc,Fvmvc} have explored speaker generation based on text prompts or face images. These studies align specific modal representations with speaker embeddings, thereby controlling the voice characteristics using the aligned representations during inference. In addition to the aforementioned absolute voice descriptions, VoxEditor \cite{voxeditor} introduces the relative descriptions for voice attributes editing, allowing for more nuanced control over voice characteristics.

\begin{figure}[t] 
  \includegraphics[width=0.5\textwidth]{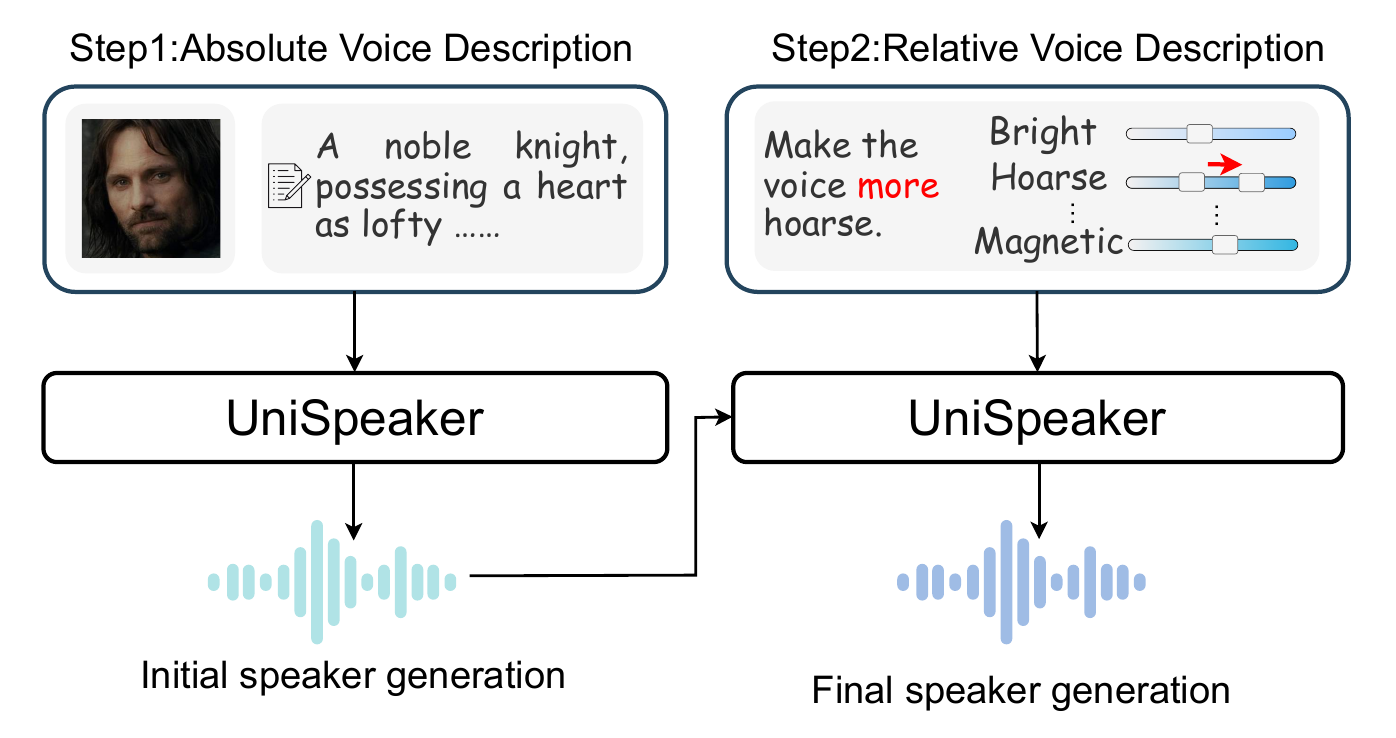}
  \caption{The pipeline of Unispeaker for multiple modalities speaker generation. Initial speaker generation is performed using absolute voice descriptions. If the initial results are unsatisfactory, further voice attributes editing can be done to achieve a finsal speaker generation.}
\label{fig1}
\end{figure}
% 而由于一对多问题的存在，
 Despite significant progress in these studies, previous methods often explore different voice description
modalities and generation approaches independently, typically involving only one extra modality aligned with the reference speech. 
 This leads to two shortcomings: (1) Independent modal alignments hindered collaborative speaker generation across multiple modal descriptions. Actually, the mapping between modalities other than the reference speech and voice characteristics is one-to-many \cite{prompttts2}, meaning the absolute voice description (a single face or a text description) can often correspond to different reasonable voice characteristics. If we combine the absolute voice description and relative voice description, it is evident that the generated speaker will better align with user needs. However, existing models can only process absolute or relative descriptions.
% If we describe the desired voice characteristics through multiple modalities, it is evident that when the model can generate voices that meet these multiple modalities, it will better align with user needs. However, existing models can only process a single input and cannot generate voices that satisfy multi-modal descriptions at once.
% Hence, it is difficult to create a voice that fully meets the user's expectations using just a single text description or an image. 
% meaning it is difficult to create a voice that fully meets the user's expectations using just a single text description or an image. 
% 这意味着一张人脸和一句文本描述往往可以对应多种合理的音色。一个简单有效的方法是，我们通过多种多维度地去描述想要的音色属性，当模型能够生成符合多种维度的音色时，这显然会让生成音色更符合用户的需求。但现有的模型只能处理一种输入，无法生成满足多模态音色描述的at once.
% when given a facial image and a speaker identity description, it was difficult to generate speech that conforms to both voice descriptions at once.  
(2)  Multimodal speech alignment data is scarce and previous methods were trained from scratch on such limited paired multimodal data. This leads to a sparse coverage of the voice space and the limited diversity and consistency of the voice characteristics generated.

% 理想且实际的音色创造流程应该是这样的，首先，最基础的，模型可以接受单一模态的绝对音色描述进行音色生成。进一步，模型可以同时接受人脸和文本两种音色描述去生成复合两种模态音色，这样用户可以通过多模态更精准地表达音色需求。最后，考虑到一对多问题，如果用户对绝对描述所生成的音色不满意，模型可以再让用户采用相对描述对生成的音色进行细致的属性编辑，最终生成满意的音色。

% 具体的，在音色创作时，除了接受单个模态的音色描述，UniSpeaker可以同时接受人脸和文本描述，来生成匹配的音色。考虑到一对多问题，当用户对利用绝对描述所生成的音色不满意时，UniSpeaker可以对已有音色进行属性的精准编辑，直至生成满意的音色。
% 这缓解的数据稀缺性导致的
% MVA的输入送入后续的生成模型进行音色控制，同时与话者编码进行对齐。考虑到不同话者间音色之间的联系，%其中，key-value vectors存储的大量的音色信息，多模态的音色描述从中检索出最相关的信息。
To address these limitations, we introduce UniSpeaker, a unified speaker generation model that integrates both absolute and relative voice descriptions. As illustrated in Figure \ref{fig1}, UniSpeaker is capable of processing  inputs from facial and textual descriptions to generate  voice characteristics that meet the expectations set by these voice descriptions. Recognizing the one-to-many issue, wherein users might find the generated voice characteristics unsatisfactory based on the aforementioned absolute descriptions, UniSpeaker enables precise voice attribute editing on the generated speech until the desired voice is attained. To achieve collaborative speaker generation, we propose a unified multimodal voice aggregator (MFA), which aligns these multimodal inputs into a coherent voice space. The MFA is based on the KV-Former architecture, a streamlined variant of the Transformer model, utilizing learnable key-value vectors to develop a shared multimodal voice space, with multimodal representations serving as queries. These key-value vectors encapsulate sufficient voice information, allowing the multimodal voice description to extract the most pertinent information. The output from the MFA is supplied to a subsequent generative model for voice control and aligned with speaker embeddings. In light of the correlation between the voice characteristics of different speakers, soft contrastive learning (SoftCL) is employed during alignment training, which relaxes strict one-to-one contrastive constraints and utilizes intra-modal discriminative information for guidance. Similar with ImageBind, this speech-anchoring mechanism facilitates the emergent alignment of various modalities within the voice space without parallel data across all modalities, which mitigated the impact of data scarcity and ensure the diversity of the voice characteristics.
In addition, large-scale speech generation models excel in voice control, but scalable multimodality integration is yet to be explored. We use the open-source CosyVoice \cite{cosyvoice} as the backbone for UniSpeaker and apply self-distillation \cite{seedtts} to enhance voice disentanglement, maintaining its versatility across tasks.

% 英文翻译，除了上述基础实验外，我们进一步在多模态同时输入时，UniSpeaker音色控制的能力，并通过主观实验验证了其有效性。
Due to the lack of publicly accessible benchmarks for assessing multimodality-driven voice control, we developed a multimodality-based voice control (MVC) benchmark. This benchmark covers five fundamental tasks: face-driven voice conversion (FaceVC), face-driven personalized text-to-speech (FaceTTS), text description-driven voice conversion (TextVC), text description-driven personalized text-to-speech (TextTTS), and attribute-driven voice editing (AVE). 
% It also includes two collaborative tasks: face and text-driven voice conversion (FaceTextVC) and face and text-driven personalized text-to-speech (FaceTextTTS). 
Consistent with prior research \cite{promptvc}, the MVC benchmark evaluates generated speech using multimodal voice descriptions on three parameters: voice suitability, voice diversity, and speech quality. We assessed UniSpeaker with the MVC benchmark, where it outperformed previous modality-specific models in the five fundamental tasks. 
% Compared to the fundamental tasks, we found that the speech generated by Unispeaker in collaborative tasks aligns more closely with user needs, demonstrating the superiority of speaker generation through multimodal collaboration.
%  相比基础任务，我们发现Unispeaker在collaborative tasks生成的语音在主观测试中更贴近用户的需求，证明了多模态协作的说话人生成的优越性。

\section{Relate Work}
\subsection{Multimodality-driven voice control for speech generation} 
Modeling diverse voice characteristics has consistently been a critical focus in the field of speech synthesis. Recent works, such as PromptTTS2 \cite{prompttts2}, Audiobox \cite{audiobox}, InstructSpeech\cite{instructspeech} and others \cite{mmtts,instructtts,textrolspeech}, have explored using text prompts to control the style or emotion of generated speech. However, only a few studies have specifically targeted voice control with text prompt \cite{prompttts++,promptspeaker}. 
Text prompt-based style control TTS methods typically convert speech attributes like pitch, energy, duration, and emotion into natural style prompts using LLMs. Since these style prompts primarily reflect prosody and capture minimal speaker individuality, achieving the desired voice control remains challenging.

In the field of multimodal voice control, researchers have previously attempted to align different voice description modalities with speaker embeddings using models such as memory networks \cite{Fvmvc}, mixture density networks \cite{prompttts++}, and latent diffusion \cite{promptvc}, as well as loss functions like MSE loss \cite{facevc}, cosine similarity loss \cite{promptspeaker}, and perceptual loss \cite{SP-FaceVC}. However, these alignment methods relied on parallel datasets and were challenging to extend directly to additional modalities. Performance-wise, previous face-based methods \cite{ImaginaryVoice} generally ensured gender accuracy but often produced incongruous voice characteristics, such as generating a youthful voice for an elderly face. Additionally, VoxEditor \cite{voxeditor} is limited to performing voice attribute editing on existing source speech, thus offering restricted voice diversity.  In response, the proposed UniSpeaker employs a unified voice aggregator to construct a shared voice space that can be easily extended to new modalities, achieving versatile and diverse voice control.

\subsection{Large speech generation models}
As speech generation systems \cite{ns,vits} have achieved remarkable levels of naturalness and robustness, recent research \cite{naturalspeech3,ditspeech} has shifted focus towards exploring novel generative models, advanced modeling objectives, and larger-scale datasets to pursue voice diversity. When integrating multimodal voice descriptions, it is crucial to preserve the performance of pre-trained speech generation models in terms of naturalness, robustness, and prosody.  Some representative large-scale speech generation \cite{valle,clamtts,valle2} models typically leverage a neural codec to convert speech waveforms into discrete acoustic token sequences, along with an autoregressive language model to generate discrete tokens from text. However, the discrete acoustic token sequences entangle content, speaker, and prosodic information in this approach, complicating the alignment of multimodal voice characteristics without disrupting the content and prosody of the generated speech. 
Recently, CosyVoice \cite{cosyvoice} has utilized supervised semantic tokens \cite{whisper} as the modeling objectives for a large language model (LLM). Subsequently, a conditional flow matching model (CFM) generates speech based on semantic tokens, speaker embeddings and mel spectrograms prompt. 
Since the semantic tokens primarily encompass content and prosodic information, the speaker information included is limited. This facilitates further voice disentanglement and the integration of multimodal voice descriptions, making CosyVoice well-suited as the backbone for the UniSpeaker model proposed in this paper. 

\begin{figure*}[t] 
  \includegraphics[width=\textwidth]{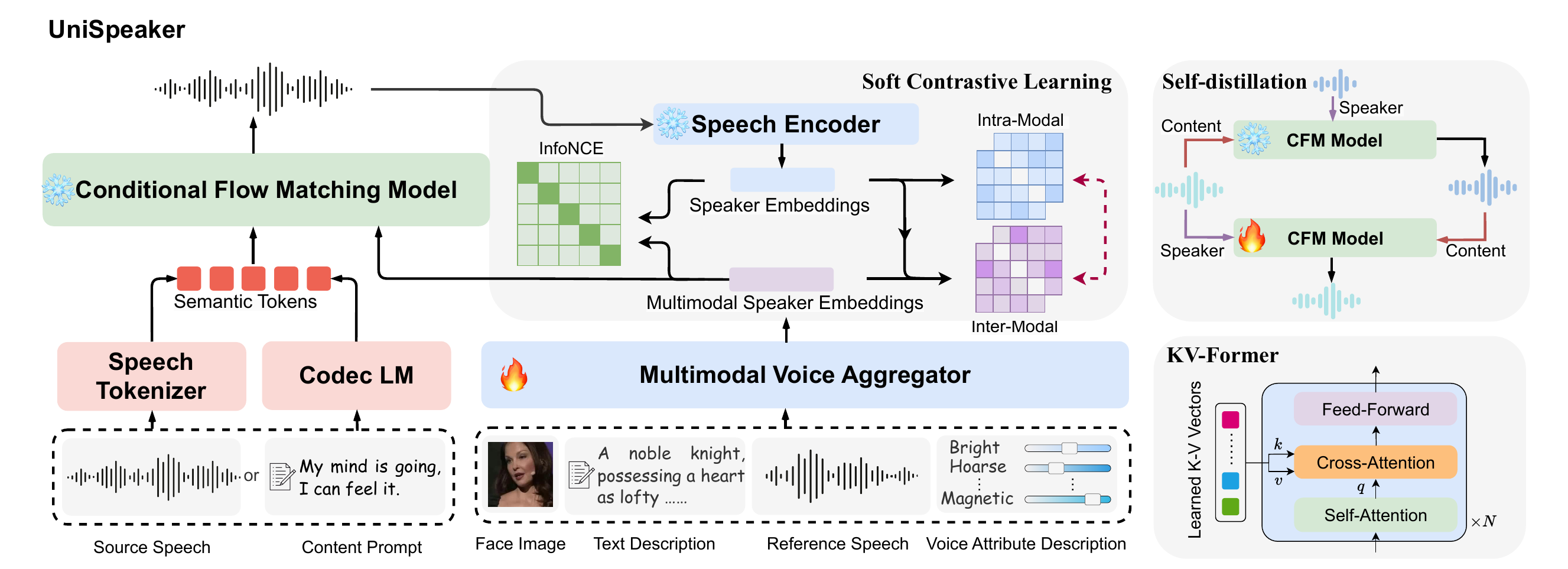}
  \caption{The pipeline of Unispeaker for multiple modalities speaker generation. Initial speaker generation is performed using absolute voice descriptions. If the initial results are unsatisfactory, further voice attributes editing can be done to achieve a finsal speaker generation.}
\label{fig1}
\end{figure*}

\section{Methods}
In this section, we first review the backbone CosyVoice, then introduce how multimodal voice descriptions are integrated into a pre-trained speech generation model.
% , and finally outline our self-distillation training strategy.

\subsection{Preliminaries}
CosyVoice leverages supervised semantic tokens \cite{whisper,asq} as modeling objectives, utilizing an LLM for text-to-token generation and a CFM for token-to-speech synthesis. 
% Here, we provide a brief overview of the LLM and CFM in CosyVoice.
Given a dataset $\mathcal{D} = \{\mathbf{x}_{i}, \mathbf{y}_{i}\}$, where $\mathbf{x}$ is a speech sample and  $\mathbf{y}$ is the corresponding text transcription, the sequence input to the LLM is mainly comprised of $\{\mathbf{s}, \mathbf{Y}, \mathbf{C}\}$, where $\mathbf{s}$ represents the speaker embeddings of $\mathbf{x}$, $\mathbf{Y}$ is the text embedding of $\mathbf{y}$ and $\mathbf{C}$ is the semantic tokens of $\mathbf{x}$. The LLM is then trained in an autoregressive manner to minimize the negative log-likelihood of semantic tokens $\mathbf{C}$.
The core of CFM is to construct a probability density path from a prior distribution to  $p_0(\mathbf{X})$ to the data distribution of the Mel-spectrograms $q(\mathbf{X})$. The probability density path is defined by a time-dependent vector field $\mathbf{v}_t(\mathbf{X})$,which generates the flow $\phi_t$ through an ordinary differential equation (ODE). The flow matching model is trained using optimal-transport conditional flow matching (OT-CFM) \cite{ot-cfm}, which can be written as follows,
\begin{equation}
	\begin{aligned}
		\mathcal{L}_\text{OT-CFM} = & \mathbb{E}_{t,p_0(\mathbf{X}_0),q(\mathbf{X}_1)} \Vert  \omega_t(\phi^{OT}_t(\mathbf{X}_0,\mathbf{X}_1)|\mathbf{X}_1) - \\ & \nu_t(\phi^{OT}_t(\mathbf{X}_0,\mathbf{X}_1)|\theta_{CFM}) \Vert_1, 
	\end{aligned}
\end{equation}
where 
% $\phi^{OT}_t(\mathbf{X}_0,\mathbf{X}_1)=(1-t)\mathbf{X}_0+t\mathbf{X}_1$ and $\omega_t(\phi^{OT}_t(\mathbf{X}_0,\mathbf{X}_1)|\mathbf{X}_1)=\mathbf{X}_1-\mathbf{X}_0.$
\begin{equation}
	\begin{aligned}
		&\phi^{OT}_t(\mathbf{X}_0,\mathbf{X}_1)=(1-t)\mathbf{X}_0+t\mathbf{X}_1, \\ &\omega_t(\phi^{OT}_t(\mathbf{X}_0,\mathbf{X}_1)|\mathbf{X}_1)=\mathbf{X}_1-\mathbf{X}_0.
	\end{aligned}
\end{equation}
The speaker embeddings $\mathbf{s}$, speech tokens $\mathbf{C}$, and masked Mel-spectrogram prompt $\tilde{\mathbf{X}}_1$ are also fed into the neural network to match the vector field with learnable parameters $\theta_{CFM}$,
\begin{equation}
\begin{aligned}
    &\nu_t(\phi^{OT}_t(\mathbf{X}_0,\mathbf{X}_1)|\theta_{CFM})=\mathrm{NN} \left(\phi^{OT}_t(\mathbf{X}_0,\mathbf{X}_1),t;\mathbf{s},\mathbf{C},\tilde{\mathbf{X}}_1\right).
\end{aligned}
\label{eq:nn}
\end{equation}
The supervised semantic tokens contains only a small amount of speaker information, and CosyVoice demonstrates good performance in voice characteristics disentanglement. In our preliminary experiments, we found that although the LLM received speaker embeddings, its impact on voice characteristics was minimal. In contrast, the CFM module plays a decisive role in influencing voice characteristics . 

\subsection{Multimodal Voice Description Integration}
We incorporate multiple modalities into the CFM model, allowing various inputs to control the voice characteristics of generated speech. As shown in Figure , each modality is first processed by a pre-trained, modality-specific encoder to obtain the corresponding representation. Each kind of representation is then transformed into a latent vector via adaptive average pooling or a multi-layer perceptron. Those vectors across modalities are mapped into a unified voice space through a shared MVA, producing the corresponding speaker embeddings. These speaker embeddings are then fed into the CFM for speech generation.

\paragraph{Multimodal Voice Aggregator} Then global representations of different modalities should be aligned with speaker embeddings within the voice space. Previous methods relied on limited datasets that matched only two modalities for alignment, resulting in a sparse distribution in the voice space and weak generalization capabilities. 
% Therefore, effectively utilizing multimodal voice descriptions through joint modeling to share a unified voice space, and thereby enhance the performance of each modality, remains an open question.

Inspired by Q-Former \cite{bilp2} and the memory mechanism \cite{Fvmvc,DBLP:conf/cvpr/LeeKCKR21}, we propose the KV-Former architecture as a unified multimodal voice aggregator. This architecture integrates learnable key-value vectors into a simplified Transformer, as shown in Figure.  The multimodal representations act as queries and perform multi-head cross-attention with the learnable key-value vectors to retrieve the most informative representation in the voice subspace. The formulation of this process is as follows,
\begin{equation}
    \mathbf{q}=\mathbf{W}^q \mathbf{s}_m, \mathbf{k}=\mathbf{W}^k \mathbf{f}, \mathbf{v}=\mathbf{W}^v \mathbf{f},
% \end{equation}
% \begin{equation}
    \mathbf{a}_{m}=\operatorname{Softmax}\left(\frac{\mathbf{q} \mathbf{k}^T}{\sqrt{d}}\right)\mathbf{v},
\end{equation}
% where  $\mathbf{W}$ are the projection matrices in attention, $ \mathbf{s}_m \in \{ \mathbf{s}_{f}, \mathbf{s}_{r}, \mathbf{s}_{t} \}$, $\mathbf{f}$ are learnable key-value vectors, $d$ is the dimension of the  $\mathbf{f}$,  $\mathbf{a}_{m}$ is the output of cross attention. 
where $\mathbf{W}$ are the projection matrices in attention, $\mathbf{s}_m \in \{ \mathbf{s}_{f}, \mathbf{s}_{r}, \mathbf{s}_{t} \}$ represents various state vectors, $\mathbf{f}$ are learnable key-value vectors, $d$ is the dimension of $\mathbf{f}$, and $\mathbf{a}_{m}$ is the output of cross attention.
In this process, the learnable key-value vectors create an information bottleneck, interacting with the three modalities to build a shared voice space. Additionally, MVA adopts a speech-anchoring mechanism, reference speech is used as input for MVA with a 50\% probability. 
In this way, even without parallel data between all modalities, different modalities achieves emergent alignment in the voice space through shared k-v vectors and joint training, which mitigated the impact of data scarcity and ensure the diversity of voice characteristics. In addition, our module also allows for easy expansion to new modalities by adding the a modality-specific encoder.

% Additionally, to ensure the voice space contains sufficient voice information, MVA adopts a speech-anchoring mechanism. Specifically, each sample includes a speech clip paired with a voice description from another modality. During training, the reference speech is used as input with a 50\% probability, enhancing the alignment of multimodal representations.

To integrate multimodal inputs for voice control without losing the general abilities of CFM, we feed the output of  MVA to the CFM and adapt the model without changing the CFM weights. The MVA is trained to optimize $\mathcal{L}_\text{OT-CFM}$ and Equation (\ref{eq:nn}) is transformed as follows to fit speaker embeddings,
\begin{equation}
	\begin{aligned}
&\nu_t(\phi^{OT}_t(\mathbf{X}_0,\mathbf{X}_1)|\theta_{MVA})=\mathrm{NN} \left(\phi^{OT}_t(\mathbf{X}_0,\mathbf{X}_1),t;\mathbf{v}_{m},\mathbf{C} \right),
	\end{aligned}
\end{equation}
where $\mathbf{v}_{m} \in \{\mathbf{v}_{f}, \mathbf{v}_{r}, \mathbf{v}_{t} \}$ and $\mathbf{v}_{f}, \mathbf{v}_{r}, \mathbf{v}_{t}$ are the outputs of applying MVA to $\mathbf{s}_{f}, \mathbf{s}_{r}, \mathbf{s}_{t}$, respectively. In this manner, CFM can integrate multiple modalities for voice control and keep its ability to generate natural and robust speech. 

\paragraph{Soft Contrastive Learning} 
Relying solely on OT-CFM to optimize MVA leads to slow convergence, and the generated speech may exhibit voice discordance with the input voice descriptions. Inspired by previous studies \cite{softclip,atr}, we additionally introduce the SoftCL strategy for speech-anchoring multimodal alignment, including both inter-modal and intra-modal alignment, as shown in Figure . For inter-modal alignment, we employ InfoNCE \cite{clip}, which pulls the paired multimodal and speaker embeddings closer together while pushing the unpaired ones apart. In addition, to bring cross-modal similarities closer to the distribution within each modality, intra-modal similarities serve as soft labels. Specifically, given a batch of $N$ multimodal-voice speaker embeddings pairs ${\{(\mathbf{v}_{m}^{i}, \mathbf{s}_{r}^{i}) \}_{i=1}^N}$, the intra-model self-similarity vector $p_{i}(\mathbf{s}_r, \mathbf{s}_{r})= \{p_{ij}(\mathbf{s}_r, \mathbf{s}_{r})\}_{j=1}^N $ can be obtained by:
\begin{equation}
    p_{i j}(\mathbf{s}_{r}, \mathbf{s}_r)=\frac{\exp \left(\operatorname{sim}\left(\mathbf{s}_{r}^{i}, \mathbf{s}_{r}^{j}\right) / \tau\right)}{\sum_{j=1}^N \exp \left(\operatorname{sim}\left(\mathbf{s}_{r}^{i}, \mathbf{s}_{r}^{j}\right) / \tau\right)},
\end{equation}
where $\tau$ is a learnable temperature coefficient, initialized to 0.07, and $\operatorname{sim}()$ denotes the dot product used to calculate similarity. Despite intra-model self-similarity, the confidence of positive samples still outweighs that of negatives, potentially overshadowing negatives in cross-modal relation alignment. To address this, we disentangle the negatives in the distribution to boost the relation alignment. For the self-similarity vector $p_{i}(\mathbf{s}_r, \mathbf{s}_{r}) \in \mathbb{R} ^ {1 \times N}$, the neg-disentangled $p_{i}^*(\mathbf{s}_r, \mathbf{s}_{r}) \in \mathbb{R} ^ {1 \times (N-1)}$  distribution is calculated as follows, 
\begin{equation}
    p_{i j}^*=\frac{\exp \left( p_{i j}\right) }{\sum_{k=1, k \neq i}^N \exp \left( p_{i k}\right)}.
\end{equation}
We also apply the above negative disentanglement to  $p_{i}(\mathbf{s}_r, \mathbf{v}_{m})$, yielding $p_{i}^{*}(\mathbf{s}_r, \mathbf{v}_{m})$.  
Then, the intra-modality alignment supervision can be achieved with negative disentanglement as follows,
% \begin{equation}
% \mathcal{L}_{\text{INTRA}}=\frac{1}{2N} \sum_{i=1}^N \mathrm{KL}\left( p_{i}^{*}(\mathbf{s}_r, \mathbf{v}_{m}) \| p_{i}^*(\mathbf{s}_r, \mathbf{s}_{r}) \right) + \frac{1}{2N} \sum_{i=1}^N \mathrm{KL}\left( p_{i}^*(\mathbf{s}_r, \mathbf{s}_{r}) \| p_{i}^{*}(\mathbf{s}_r, \mathbf{v}_{m}) \right),
% \end{equation}
\begin{equation}
\mathcal{L}_{\text{INTRA}}=\frac{1}{N} \sum_{i=1}^N \mathrm{KL}\left(p_{i}^*(\mathbf{s}_r, \mathbf{s}_{r}) \| p_{i}^{*}(\mathbf{s}_r, \mathbf{v}_{m}) \right),
\end{equation}
% 综上，UniSpeaker的训练损失如下
where $\mathrm{KL}$ represents the Kullback-Leibler Divergence. Generally, UniSpeaker is trained to optimize the following loss function,
\begin{equation}
\mathcal{L} = \mathcal{L}_\text{OT-CFM} + \lambda_{1} \mathcal{L}_{\text{INTRA}} + \lambda_{2} \mathcal{L}_{\text{INTER}},
\label{eq:total}
\end{equation}
where $\mathcal{L}_{\text{INTRA}}$ is the InfoNCE loss, $\lambda_{1}$ and $\lambda_{2}$ are hyper-parameters used to balance each loss term.

\paragraph{Self-distillation}
In our preliminary experiments, we observed that CFM usually draws speaker information from semantic tokens, often overlooking speaker information within the face image, due to the cross-modal gap. Therefore, to enhance voice disentanglement before merging multimodal voice descriptions, self-distillation is applied to fine-tune the CFM. Initially, we employ semantic tokens from the original speech, along with a Mel-spectrogram prompt and speaker embeddings from a randomly chosen speaker, which are then inputted into the CFM for voice conversion.
Then, given the semantic tokens $\bar{\mathrm{C}}$ of converted speech  and speaker embeddings $\mathbf{s}$  of source speech, the CFM is fine-tuned to predict the source speech.
We removed the masked Mel-spectrogram prompt to improve the voice control by the speaker embeddings, transforming Equation (\ref{eq:nn})  as follows,
\begin{equation}
	\begin{aligned}
&\nu_t(\phi^{OT}_t(\mathbf{X}_0,\mathbf{X}_1)|\theta_{FM})=\mathrm{NN} \left(\phi^{OT}_t(\mathbf{X}_0,\mathbf{X}_1),t;\mathbf{s},\bar{\mathrm{C}} \right).
	\end{aligned}
\end{equation}
In this way, the voice characteristics of the generated speech is controlled by the speaker embeddings input to the CFM. This allows the integration of multimodal voice description directly into the CFM, simplifying the process without requiring modifications to the LLM.

% 受监督的语义表征仅少量话者信息，CosyVoice有着很好的话说话人解耦的性能。在我们的预先实验中发现，LLM尽管接受了话者编码，但对音色的影响很小，而CFM模块对音色起着决定性的作用，因此我们在CFM模块进行了多模态表征的嵌入来控制音色。

% UniSpeaker的整体架构如图1所示，Speaker
% As illustrated in Figure 1,
% speaker embeddings and semantic tokens serve as key representations to generate speech. Speaker
% embeddings control the voice characteristics and can be extracted from various inputs. Semantic
% tokens convey the content and prosody of the generated speech, derived from either source speech
% or content prompt.

\begin{table*}[t]
\centering
\caption{ Objective and subjective evaluation results of comparison systems. The definitions of all metrics can be found in Section \ref{MVC Benchmark}. ``-'' denotes the results are not available.}
\label{SOTA_compare}
%\resizebox{73mm}{25mm}{
\resizebox{0.8\textwidth}{!}{
    \begin{tabular}{clcccccc}
    \toprule
    % \multirow{2}{*}{Task}&\multirow{2}{*}{Methods} &\multicolumn{4}{c}{Objective Evaluation} &\multicolumn{2}{c}{Subjective Evaluation}\\
    % \cmidrule(r){3-6} \cmidrule(r){6-8}
    \multirow{2}{*}{Task}&\multirow{2}{*}{Methods} &\multicolumn{3}{c}{Voice Suitability} &\multicolumn{1}{c}{Voice Diversity}  &\multicolumn{2}{c}{Speech Quality}\\
    \cmidrule(r){3-5} \cmidrule(r){6-6} \cmidrule(r){7-8}
    && SST $\uparrow$ & SSC $\uparrow$ & MOS-Match $\uparrow$ & SSD $\downarrow$ & WER $\downarrow$  & MOS-Nat $\uparrow$ \\
    \midrule
    \multirow{4}{*}{FaceTTS} &Imaginary Voice\cite{ImaginaryVoice}  & 10.08 & 38.46 &2.39 $\pm$ 0.09 & 32.17  & 8.23 & 2.45 $\pm$ 0.08  \\
    &Face-StyleSpeech\cite{face_style_tts} &11.02 & 37.09 &2.78 $\pm$ 0.12  & 30.78  & 7.09 & 3.29 $\pm$ 0.10 \\ 
    &SYNTHE-SEES\cite{SYNTHE-SEES} &10.97 & 38.81 & 2.92 $\pm$ 0.11 &31.09  & 9.14 & 3.39 $\pm$ 0.09 \\
    &UniSpeaker(Ours) &\textbf{12.48} & \textbf{40.75} &\textbf{3.18} $\pm$ \textbf{0.10} & \textbf{14.09}   &\textbf{4.01} & \textbf{3.82} $\pm$ \textbf{0.08}  \\
    \midrule
    \multirow{4}{*}{FaceVC} &FaceVC\cite{facevc}  & 8.97 & 50.91 &2.21 $\pm$ 0.11  & 30.19  & 10.90 &2.79 $\pm$ 0.10   \\
    &SP-FaceVC\cite{SP-FaceVC} & 9.52 &52.29 &2.39  $\pm$ 0.09 &29.86  & 14.92 & 3.04 $\pm$ 0.10 \\ 
    &FVMVC\cite{Fvmvc} & 9.49 &51.33 & 2.69 $\pm$ 0.07 &22.60  & 11.94 &3.31 $\pm$ 0.08  \\
    &UniSpeaker(Ours) &\textbf{11.68}  &\textbf{55.13} &\textbf{3.09} $\pm$ \textbf{0.10} &\textbf{15.91}  &\textbf{4.98} &\textbf{3.80} $\pm$ \textbf{0.09} \\
    \midrule
    \multirow{4}{*}{TextTTS} &PromptSpeaker\cite{promptspeaker}  & 17.39 &- & 3.64 $\pm$ 0.13 & 29.84   & 14.70 &3.37 $\pm$ 0.10 \\ 
    &Prompttts++\cite{prompttts++} & 16.87 &- &3.63 $\pm$ 0.12  & 35.42   & 15.08 &3.41 $\pm$ 0.11 \\ 
    &CosyVoice-Instruct \cite{cosyvoice} &14.51 &- &3.71 $\pm$ 0.13 &34.62  &7.03 & \textbf{3.91} $\pm$ \textbf{0.09} \\
    &UniSpeaker (Ours) &\textbf{23.09}	&- &\textbf{3.85} $\pm$ \textbf{0.11} &\textbf{21.10}  & \textbf{6.46} &3.87 $\pm$ 0.13 \\
    \midrule 	
    \multirow{2}{*}{TextVC} &PromptVC\cite{promptvc}  & 16.59	&- &3.47 $\pm$ 0.07  & 36.98  & 7.08 &3.64 $\pm$ 0.10 \\
    &UniSpeaker(Ours) &\textbf{24.45}  &- &\textbf{3.81} $\pm$ \textbf{0.09}  &\textbf{24.04} & \textbf{6.29}  &\textbf{3.77} $\pm$ \textbf{0.11}   \\
    \midrule
    \multirow{2}{*}{AVE} &VoxEditor\cite{voxeditor}  &41.48 &-	& \textbf{3.78} $\pm$ \textbf{0.09} &49.92  & 8.01  & 3.57 $\pm$ 0.10 \\
    &UniSpeaker(Ours) &\textbf{49.04} 	&- &\textbf{3.79} $\pm$ \textbf{0.10} &\textbf{34.92}  & \textbf{4.09} & \textbf{3.92} $\pm$ \textbf{0.09} \\
    \bottomrule
    \end{tabular}
}
\end{table*}

\section{Dataset and Benchmark}
\label{MVC Benchmark}
Four modality-specific datasets were used to train the UniSpeaker, including LRS3-TED \cite{lrs3_ted}, LibriTTS-P \cite{libritts-p}, VCTK-R \cite{voxeditor}, and inner speaker identity description dataset collected from the internet, totaling about 1000 hours of audio data..

In the MVC Benchmark, for face-related evaluation, we randomly selected 600 face images from the test set of LRS3-TED. In terms of textual descriptions, 600 sentences were randomly picked from the validation set and rewritten by a LLM (GPT-3.5-TURBO), ensuring that the meaning of the sentences remained unchanged. For voice attribute editing, 200 sentences were randomly selected from VCTK and edited on all attributes for evaluation. All above samples are unseen during training. The MVC benchmark evaluates the generated speech from three perspectives: voice suitability, voice diversity, and speech quality. 1) \textbf{Voice suitability} evaluates whether the voice characteristics of the generated speech align with the input multimodal voice description. This includes three specific metrics: Speaker Similarity with Target (SST), Speaker Similarity Consistency (SSC), and MOS-Match. 
% Speaker similarity can be computed by cosine similarity between the speaker embeddings, which are extracted from speech using a speaker verification model\footnote{\url{https://github.com/modelscope/3D-Speaker}}. SST is measured by calculating the speaker similarity between the generated speech and reference speech of the target speaker. SSC assesses the consistency of the generated voice with various descriptions for the same speaker by calculating speaker similarity between the speech generated from different face images of the same speaker. 
% MOS-Match is obtained through subjective listening tests for the mean opinion score to quantify how closely the voice characteristics of the generated speech align with the input description.
2) \textbf{Voice diversity} evaluates the model's ability to produce a diverse set of voice characteristics based on the descriptions of different speakers, rather than generating very similar ones. A metric named Speaker Similarity Diversity (SSD) is employed for evaluating voice diversity, which measures the speaker similarity between the speech generated from the descriptions of different speakers.
3) \textbf{Speech quality} assesses the robustness and naturalness of the generated speech, using two key metrics: word error rate (WER) and MOS-Nat. We employ an automatic speech recognition model\footnote{\url{https://huggingface.co/openai/whisper-large-v3}}  to transcribe the generated speech and compute the WER. MOS-Nat is determined through subjective listening tests for mean opinion scores to evaluate the naturalness of the generated speech. Please refer to Appendix for more details.

\section{Experiments}
\subsection{Experiment Settings}
% LLM,CFM的模型参数和CosyVoice一致。MVA的模型参数见表7。当进行text-to-speech任务时，LLM只接受文本输入，不接受话者编码输入。为了与过去的方法对比，我们在实验中只接受了一种相应的输入来控制音色。在后续讨论中，我们也展示了模型同时接受多模态输入来进行更多样化的音色创造。

We trained the UniSpeaker using 4 NVIDIA TESLA V100 32G GPUs for 30K steps. The models were optimized using the AdamW optimizer with a learning rate of 1e-5 and a 10K warmup steps. The weights $\lambda_{1}$ and $\lambda_{2}$ in Equation (\ref{eq:total}) were set to 0.05. FaceNet\cite{facenet}, T5\cite{t5}, and CAM++\cite{cam++} serve as modality-specific encoders for face, text, and speech, respectively. The speech tokenizer and codec LM were the same as those used in CosyVoice. For TTS, the codec LM accepted only text inputs without speaker embeddings.  We compared our UniSpeaker with 11 task-specific expert models in five tasks. We used the official code or pre-trained checkpoints of Imaginary Voice \cite{ImaginaryVoice}, FaceVC \cite{facevc}, SP-FaceVC \cite{SP-FaceVC}, FVMVC \cite{Fvmvc}, and CosyVoice-Instruct \cite{cosyvoice}. For the other methods, we reproduced them according to their respective papers and evaluated them on the same dataset. Please refer to Appendix for more details. 

% For comparison with previous methods, only a single corresponding voice description modality serves as input.
% Details of model training and inference are provided in Appendix \ref{apx:model_training_and_inference}. 

% with 95% confidence intervals.
\begin{figure}[t] 
  \includegraphics[width=0.5\textwidth]{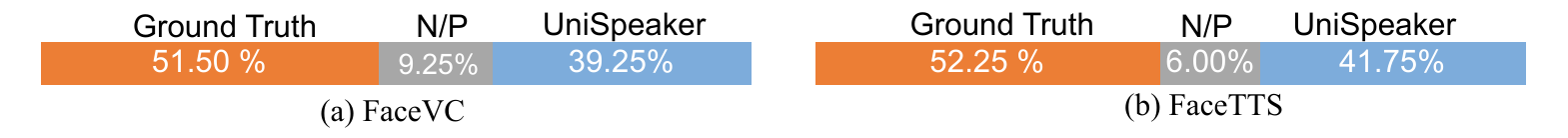}
  \caption{Average preference scores (\%) of ABX tests about voice suitability in comparison, where ``N/P'' stands for “no preference”.}
\label{pic:abx}
\end{figure} 

% 下面，我们主要从音色合理性，音色多样性以及自然度三方面来.
\subsection{Evaluation Results}
In this section, we conduct experiments comparing the UniSpeaker with the baselines and all objective and subjective evaluation results are reported in Table \ref{SOTA_compare}. 
% 我们利用ssim，ssc，mos-vox来评估音色的合理性。我们发现，（1）在5项任务上，我们的方法在上述三个指标均优于过去的方法，除了mox-vox的评分在vae任务上与baseline是可比的。（2）在人脸控制的音色的效果上，过去方法基本上只控制了音色的性别的准确的，存在许多年龄等主观感觉上的音色不合理。而我们的方法在年龄和主观感觉上有了很大的改进。此外，我们还进行了abx实验，即给一句根据参考话者人脸所生成的语音和该说话人的真实录音，要求听着选一句哪一句的音色和人脸更匹配，实验结果如图3所示，当输入人脸清晰时, 我们的方法在生成语音根据某些人脸生成的语音甚至比真实说话人的音色更匹配。(3) 在文本控制方面，CosyVoice-instruct，将说话人文本prompt拼接在内容prompt前面，在数据集有限的情况下，由于没有预训练好的文本prompt提取器，他的泛化能力很受限。（4）在音色编辑任务上，voxeditor引入了复杂的残差记忆网络，但我们的统一可扩展的说话人压缩器的性能依然是可比的。

In terms of voice suitability, our findings revealed that: 1) Across five tasks, UniSpeaker outperformed previous approaches on all three metrics, except for MOS-Match in the AVE task. While VoxEditor incorporates a complex residual memory network, the performance of our unified and scalable MVA remains competitive in MOS-Match.  2) In terms of face-based voice control, previous methods were generally effective in accurately controlling the gender of the voice characteristics but often exhibited obvious voice inconsistencies in subjective aspects such as age. In contrast, UniSpeaker achieved substantial improvements in both voice-age matching and overall subjective perception. 3) Additionally, we conducted an ABX test, as shown in Figure \ref{pic:abx}, the voice characteristics generated by UniSpeaker sometimes can match the face image even more closely than those of the actual speaker.  We encourage readers to listen to the samples on the demo page. 
4) In text control, CosyVoice-instruct concatenates voice characteristic descriptions with the content prompt in the LLM without utilizing a pre-trained text prompt, resulting in difficulties grasping semantic information effectively and producing ambiguous voice characteristics.  In contrast, UniSpeaker achieves excellent semantic-to-voice consistency, where similar semantics generate similar voice characteristics. 

% However, with limited datasets, absence of a pre-trained text prompt extractor and , its generalization capability is significantly constrained. 
% 这导致，CosyVoice 不能很好的理解语义信息，生成的音色属性很模棱两可。而UniSpeaker通过预训练好的T5模型，能有很好的语义-音色分布一致性， 即相近的语义生成相似的音色。

\begin{figure}[t] 
  \includegraphics[width=0.5\textwidth]{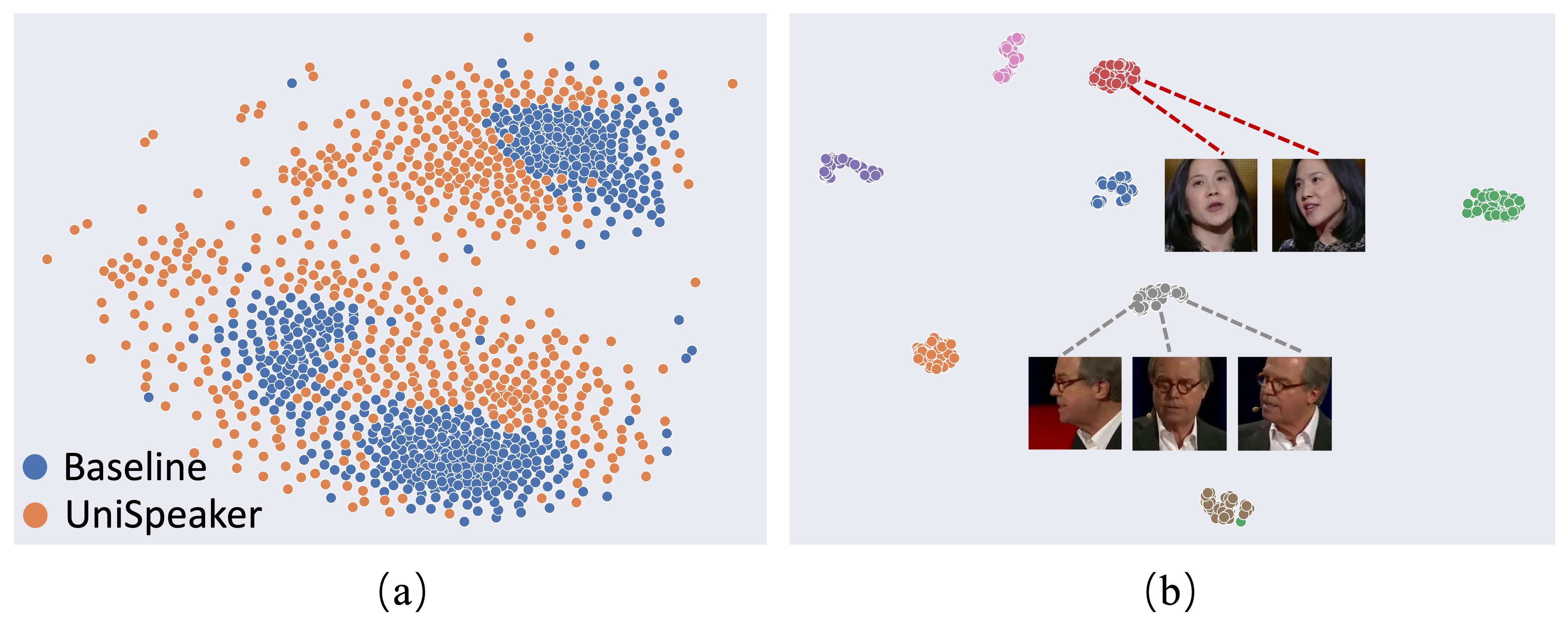} 
  \caption{The evaluation results about different multimodal data scales on joint voice modeling}
\label{diversity}
\end{figure}

% 我们用ssd 指标来评估音色的多样性，即不同的音色描述输入应该能生成多样的音色。我们可以看到，相比过去的方法，UniSpeaker 在音色多样性上显然优于过去的方法。进一步，我们从SYNTHE_SEES和UniSpeaker中随机选了600句生成的语音，他们的音色由人脸控制，用tsne对生成语音的话者编码进行了可视化，如图4所示。从图中，我们看到过去方法产生的音色空间的分布是很不均匀的，不同的人脸的音色会非常相似，导致音色的多样性非常受限。
% \paragraph{Voice Diversity:} 
In terms of voice diversity, it is clear that UniSpeaker significantly outperforms previous methods across 5 tasks.  Furthermore, we visualized the speaker embeddings of the generated speech from both SYNTHE-SEES and UniSpeaker systems using t-SNE \cite{tsne}, as shown in Figure \ref{diversity} (a). The figure reveals that the voice space generated by our
method is significantly richer, whereas the voice space of the baseline is relatively sparse. This indicates the voice characteristics generated by the baseline for different faces may being very similar, greatly limiting voice diversity.
% 接着，我们利用wer和mos-sn两个指标来评估生成语音的质量。由于在训练时，我们冻住了flow，是的UniSpeaker继承了经过self-蒸馏的CosyVoice的语音生成能力。因此，相比过去的方法，unispeaker生成语音的质量均好于过去的方法，除了在mos-nat上逊于CosyVoice-Instract。这是由于UniS peaker在某些文本描述的情况下，会学到数据集中的特定噪音，反而影响了音质。

In terms of speech quality, by freezing the CFM during training, UniSpeaker preserve the general abilities of our backbone. Consequently, UniSpeaker surpasses previous methods in overall speech quality, only the MOS-Nat slightly lags behind CosyVoice-Instruct. This lag is due to the CFM  occasionally learning noise patterns from the dataset. Conversely, CosyVoice-Instruct only integrate multimodal voice descriptions in the LLM, resulting in minimal impact on speech quality.
% 然而 CosyVoice-instruct只在LLMb部分进行嵌入，对音质影响很小。
 % 为了验证自蒸馏的有效性，1）我们对比了没有经过自蒸馏的模型，即开源的CosyVoice模型和UniSpeaker在TTS和VC任务上，根据参考语音来控制音色的性能。2）为了验证MVA的有效性，我们移除了MVA, 将模态特定编码器的输出映射到一个全局表征后，直接送入cfm. 3)为了验证softcl的有效性，我们对mva的输出，移除了类内和类间的对比损失约束。
% 1)当我们移除self-distillation时，原本的CosyVoice在进行tts任务时，llm部分仍需要话者编码，但我们的unispeaker在llm没有加入话者编码，仅依赖cfm进行话者话者。实验结果表征，自蒸馏对话者控制在对音色相似性，音色一致性上和mos-vox有了明显的改善。但由于自蒸馏所用数据的有限，在音色多样性方面造成了cosyvoice的模型遗忘。2）MVA 对模型的在音色控制上的有益的，关于MVA超参数以及纯语音数据的数量对任务的影响可以见附录。3）删除softcl对也会降低各个指标。实际上，删除softcl会导致生成音色与输入的音色描述很不匹配。

\subsection{Ablation Study}
Three ablation studies were conducted in our experiments. 1) To verify the effectiveness of MVA, the output of modality-specific encoders was mapped to the global representation, and it was directly fed into the CFM. 2) To assess the effectiveness of SoftCL, we removed the intra-class and inter-class contrastive losses from the output of MVA. 3) To validate the effectiveness of self-distillation, the performance of UniSpeaker and the open-source CosyVoice model (without self-distillation) was compared on TTS and VC tasks. We report the evaluation results for certain tasks in Table \ref{tab: ablation}, with more evaluation results available in the Appendix.

% MAV证明对音色控制是有益的，他利用更多的语音数据来共享音色空间，使得音色空间分布更均匀，从而提升了性能。
% 他利用了更多的多模态数据通过共享的k-v vectors来进行联合建模，使得音色空间分布更均匀，促进了不同模态之间的对齐，在音色多样性和音色合适性两方面提升了模型性能。
We have the following observations: 1) MVA proved beneficial for voice control with a shared multimodal voice space. It utilizes multimodal data for joint modeling through shared k-v vectors, resulting in a uniform distribution of the voice space. This promotes alignment between different modalities and enhances the model's performance in both voice diversity and voice suitability. 2) Removing SoftCL resulted in a decline across various metrics, specifically creating a significant mismatch between the generated voice and the input voice descriptions. 3) Eliminating self-distillation also had notable effects. Experimental results indicated that self-distillation significantly enhanced voice control, particularly in terms of SST. However, due to the limited data used for self-distillation, there was a slight reduction in voice diversity.

\begin{table}[t]
\centering
\caption{The ablation study of UniSpeaker, measured by SST, SSD and SSC.}
\resizebox{0.38\textwidth}{!}{
    \begin{tabular}{clccc}
    \toprule
    Task & Methods & SST $\uparrow$ & SSD $\downarrow$ & SSC $\uparrow$\\
    \midrule
    \multirow{3}{*}{FaceTTS} &UniSpeaker  &12.48 &14.09 &40.75 \\
    &w.o. MVA  &11.40	&15.07	&40.61 \\ 
    &w.o. SoftCL  &11.57	&15.94	&38.28\\
    \midrule
    \multirow{3}{*}{FaceVC} &UniSpeaker  &11.68	&15.91	&55.13 \\
    &w.o. MVA  &10.70	&19.07	&54.61 \\ 
    &w.o. SoftCL &11.08	 &19.24	&51.55 \\
    \midrule
    \multirow{2}{*}{TTS} &UniSpeaker  &44.30 &10.03 & 33.32 \\
    &w.o. self-distillation &38.49 &9.80 &29.68\\
    \midrule
    \multirow{2}{*}{VC} &UniSpeaker  &39.37	&10.34	&50.64 \\
    &w.o. self-distillation &31.07	&10.16	&43.62 \\
    \bottomrule
    \end{tabular}    
}
\label{tab: ablation}
\end{table}

\begin{figure}[t] 
  \includegraphics[scale=0.23]{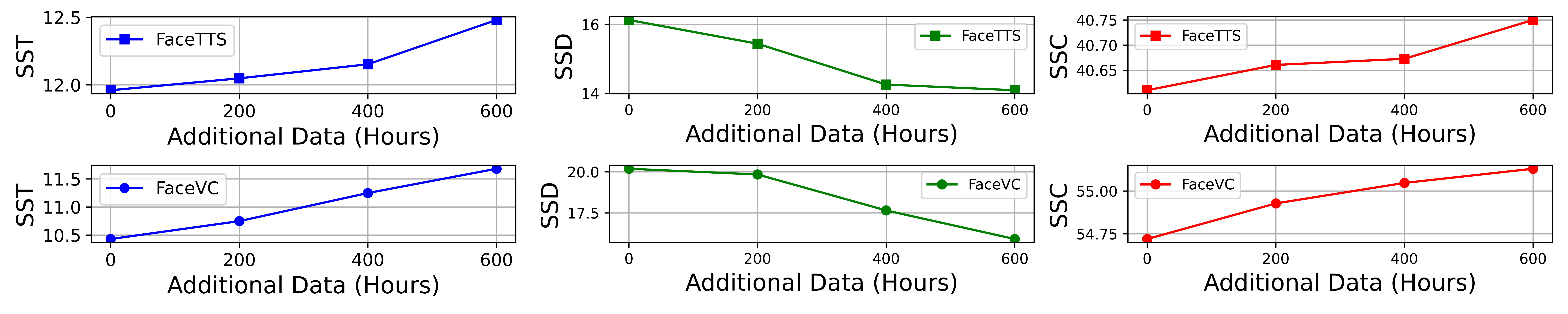}
    \caption{The evaluation results about different multimodal data scales on joint voice modeling. Here, the horizontal axis represents the amount of additional multimodal data, with ``0'' indicating that only the LRS3 dataset was used.}
\label{pic:dis1}
\end{figure} 

\subsection{Discussions}
\label{sec:dis}
% 
% \paragraph{}% 我们进一步不同数据规模对音色联合建模的影响。具体的，对人脸控制音色实验，我们分别只在人脸数据集lrs3以及加上不同规模的其他模态数据集上进行实验，其实验结果如图3所示。 从图中，我们可以看到，随着其他模态数据的增加，FaceVC和FaceTTS的性能也会得到提升，这验证了多模态联合建模的优势，有助于跨模态对齐和音色空间更丰富。相比SST和SSD, 其他模态的数据量对音色一致性的影响不大，这是因为音色一致性更注重模态内的联系。
We investigated the impact of different multimodal data scales on the shared voice space.  For face-driven voice control, we trained UniSpeaker using various datasets: solely LRS3, and additional datasets of varying sizes. The results, presented in Figure \ref{pic:dis1},  show that increasing the amount of multimodal data improves the performance of FaceVC and FaceTTS, highlighting the benefits of multimodal joint modeling. Furthermore, the influence of additional multimodal data on SSC is less pronounced for SST and SSD, as SSC primarily relies on intra-modal relationships.
We randomly selected 8 unseen speakers and sampled 100 different face images from each for FaceTTS. The t-SNE visualization of speaker embeddings extracted from generated speech is presented in Figure \ref{diversity} (b). We observed that for each speaker, the voice remained consistent across various facial images with different angles and backgrounds. This indicates that UniSpeaker demonstrates strong robustness to noisy information in facial images.

\section{CONCLUSION}
In this paper, we propose the UniSpeaker, a speech generation model that leverages multimodal voice description for voice control. Through a unified voice aggregator and  designed training strategies, UniSpeaker outperforms previous modality-specific models across five tasks, generating voices that better match the input voice descriptions. In the future, we will explore how to more effectively utilize multiple voice descriptions of different modalities for one speaker simultaneously and apply our method on other more modalities for voice control.

\bibliographystyle{named}
\bibliography{ijcai25}

\begin{thebibliography}{}

\bibitem[\protect\citeauthoryear{Afouras \bgroup \em et al.\egroup }{2018}]{lrs3_ted}
Triantafyllos Afouras, Joon~Son Chung, and Andrew Zisserman.
\newblock {LRS3-TED:} a large-scale dataset for visual speech recognition.
\newblock {\em CoRR}, abs/1809.00496, 2018.

\bibitem[\protect\citeauthoryear{Anastassiou \bgroup \em et al.\egroup }{2024}]{seedtts}
Philip Anastassiou, Jiawei Chen, Jitong Chen, Yuanzhe Chen, Zhuo Chen, Ziyi Chen, Jian Cong, Lelai Deng, Chuang Ding, Lu~Gao, Mingqing Gong, Peisong Huang, Qingqing Huang, Zhiying Huang, Yuanyuan Huo, Dongya Jia, Chumin Li, Feiya Li, Hui Li, Jiaxin Li, Xiaoyang Li, Xingxing Li, Lin Liu, Shouda Liu, Sichao Liu, Xudong Liu, Yuchen Liu, Zhengxi Liu, Lu~Lu, Junjie Pan, Xin Wang, Yuping Wang, Yuxuan Wang, Zhen Wei, Jian Wu, Chao Yao, Yifeng Yang, Yuanhao Yi, Junteng Zhang, Qidi Zhang, Shuo Zhang, Wenjie Zhang, Yang Zhang, Zilin Zhao, Dejian Zhong, and Xiaobin Zhuang.
\newblock Seed-tts: {A} family of high-quality versatile speech generation models.
\newblock {\em CoRR}, abs/2406.02430, 2024.

\bibitem[\protect\citeauthoryear{Chan \bgroup \em et al.\egroup }{2019}]{tsne}
David~M. Chan, Roshan Rao, Forrest Huang, and John~F. Canny.
\newblock {GPU} accelerated t-distributed stochastic neighbor embedding.
\newblock {\em J. Parallel Distributed Comput.}, 131:1--13, 2019.

\bibitem[\protect\citeauthoryear{Chen \bgroup \em et al.\egroup }{2024}]{valle2}
Sanyuan Chen, Shujie Liu, Long Zhou, Yanqing Liu, Xu~Tan, Jinyu Li, Sheng Zhao, Yao Qian, and Furu Wei.
\newblock {VALL-E} 2: Neural codec language models are human parity zero-shot text to speech synthesizers.
\newblock {\em CoRR}, abs/2406.05370, 2024.

\bibitem[\protect\citeauthoryear{Du \bgroup \em et al.\egroup }{2024}]{cosyvoice}
Zhihao Du, Qian Chen, Shiliang Zhang, Kai Hu, Heng Lu, Yexin Yang, Hangrui Hu, Siqi Zheng, Yue Gu, Ziyang Ma, Zhifu Gao, and Zhijie Yan.
\newblock Cosyvoice: {A} scalable multilingual zero-shot text-to-speech synthesizer based on supervised semantic tokens.
\newblock {\em CoRR}, abs/2407.05407, 2024.

\bibitem[\protect\citeauthoryear{Gao \bgroup \em et al.\egroup }{2024}]{softclip}
Yuting Gao, Jinfeng Liu, Zihan Xu, Tong Wu, Enwei Zhang, Ke~Li, Jie Yang, Wei Liu, and Xing Sun.
\newblock Softclip: Softer cross-modal alignment makes {CLIP} stronger.
\newblock In Michael~J. Wooldridge, Jennifer~G. Dy, and Sriraam Natarajan, editors, {\em Thirty-Eighth {AAAI} Conference on Artificial Intelligence, {AAAI} 2024, Thirty-Sixth Conference on Innovative Applications of Artificial Intelligence, {IAAI} 2024, Fourteenth Symposium on Educational Advances in Artificial Intelligence, {EAAI} 2014, February 20-27, 2024, Vancouver, Canada}, pages 1860--1868. {AAAI} Press, 2024.

\bibitem[\protect\citeauthoryear{Guan \bgroup \em et al.\egroup }{2024}]{mmtts}
Wenhao Guan, Yishuang Li, Tao Li, Hukai Huang, Feng Wang, Jiayan Lin, Lingyan Huang, Lin Li, and Qingyang Hong.
\newblock {MM-TTS:} multi-modal prompt based style transfer for expressive text-to-speech synthesis.
\newblock In Michael~J. Wooldridge, Jennifer~G. Dy, and Sriraam Natarajan, editors, {\em Thirty-Eighth {AAAI} Conference on Artificial Intelligence, {AAAI} 2024, Thirty-Sixth Conference on Innovative Applications of Artificial Intelligence, {IAAI} 2024, Fourteenth Symposium on Educational Advances in Artificial Intelligence, {EAAI} 2014, February 20-27, 2024, Vancouver, Canada}, pages 18117--18125. {AAAI} Press, 2024.

\bibitem[\protect\citeauthoryear{Guo \bgroup \em et al.\egroup }{2023}]{prompttts}
Zhifang Guo, Yichong Leng, Yihan Wu, Sheng Zhao, and Xu~Tan.
\newblock Prompttts: Controllable text-to-speech with text descriptions.
\newblock In {\em {IEEE} International Conference on Acoustics, Speech and Signal Processing {ICASSP} 2023, Rhodes Island, Greece, June 4-10, 2023}, pages 1--5. {IEEE}, 2023.

\bibitem[\protect\citeauthoryear{Huang \bgroup \em et al.\egroup }{2024}]{instructspeech}
Rongjie Huang, Ruofan Hu, Yongqi Wang, Zehan Wang, Xize Cheng, Ziyue Jiang, Zhenhui Ye, Dongchao Yang, Luping Liu, Peng Gao, et~al.
\newblock Instructspeech: Following speech editing instructions via large language models.
\newblock In {\em Forty-first International Conference on Machine Learning}, 2024.

\bibitem[\protect\citeauthoryear{Ji \bgroup \em et al.\egroup }{2024}]{textrolspeech}
Shengpeng Ji, Jialong Zuo, Minghui Fang, Ziyue Jiang, Feiyang Chen, Xinyu Duan, Baoxing Huai, and Zhou Zhao.
\newblock Textrolspeech: {A} text style control speech corpus with codec language text-to-speech models.
\newblock In {\em {IEEE} International Conference on Acoustics, Speech and Signal Processing, {ICASSP} 2024, Seoul, Republic of Korea, April 14-19, 2024}, pages 10301--10305. {IEEE}, 2024.

\bibitem[\protect\citeauthoryear{Ju \bgroup \em et al.\egroup }{2024}]{naturalspeech3}
Zeqian Ju, Yuancheng Wang, Kai Shen, Xu~Tan, Detai Xin, Dongchao Yang, Eric Liu, Yichong Leng, Kaitao Song, Siliang Tang, Zhizheng Wu, Tao Qin, Xiangyang Li, Wei Ye, Shikun Zhang, Jiang Bian, Lei He, Jinyu Li, and Sheng Zhao.
\newblock Naturalspeech 3: Zero-shot speech synthesis with factorized codec and diffusion models.
\newblock In {\em Forty-first International Conference on Machine Learning, {ICML} 2024, Vienna, Austria, July 21-27, 2024}. OpenReview.net, 2024.

\bibitem[\protect\citeauthoryear{Kang \bgroup \em et al.\egroup }{2023}]{face_style_tts}
Minki Kang, Wooseok Han, and Eunho Yang.
\newblock Face-stylespeech: Improved face-to-voice latent mapping for natural zero-shot speech synthesis from a face image.
\newblock {\em CoRR}, abs/2311.05844, 2023.

\bibitem[\protect\citeauthoryear{Kawamura \bgroup \em et al.\egroup }{2024}]{libritts-p}
Masaya Kawamura, Ryuichi Yamamoto, Yuma Shirahata, Takuya Hasumi, and Kentaro Tachibana.
\newblock Libritts-p: {A} corpus with speaking style and speaker identity prompts for text-to-speech and style captioning.
\newblock {\em CoRR}, abs/2406.07969, 2024.

\bibitem[\protect\citeauthoryear{Kim \bgroup \em et al.\egroup }{2021}]{vits}
Jaehyeon Kim, Jungil Kong, and Juhee Son.
\newblock Conditional variational autoencoder with adversarial learning for end-to-end text-to-speech.
\newblock In Marina Meila and Tong Zhang, editors, {\em Proceedings of the 38th International Conference on Machine Learning, {ICML} 2021, 18-24 July 2021, Virtual Event}, volume 139 of {\em Proceedings of Machine Learning Research}, pages 5530--5540. {PMLR}, 2021.

\bibitem[\protect\citeauthoryear{Kim \bgroup \em et al.\egroup }{2024}]{clamtts}
Jaehyeon Kim, Keon Lee, Seungjun Chung, and Jaewoong Cho.
\newblock Clam-tts: Improving neural codec language model for zero-shot text-to-speech.
\newblock In {\em The Twelfth International Conference on Learning Representations, {ICLR} 2024, Vienna, Austria, May 7-11, 2024}. OpenReview.net, 2024.

\bibitem[\protect\citeauthoryear{Lee \bgroup \em et al.\egroup }{2021}]{DBLP:conf/cvpr/LeeKCKR21}
Sangmin Lee, Hak~Gu Kim, Dae~Hwi Choi, Hyung{-}Il Kim, and Yong~Man Ro.
\newblock Video prediction recalling long-term motion context via memory alignment learning.
\newblock In {\em {IEEE} Conference on Computer Vision and Pattern Recognition, {CVPR} 2021, virtual, June 19-25, 2021}, pages 3054--3063. Computer Vision Foundation / {IEEE}, 2021.

\bibitem[\protect\citeauthoryear{Lee \bgroup \em et al.\egroup }{2023}]{ImaginaryVoice}
Jiyoung Lee, Joon~Son Chung, and Soo{-}Whan Chung.
\newblock Imaginary voice: Face-styled diffusion model for text-to-speech.
\newblock In {\em {IEEE} International Conference on Acoustics, Speech and Signal Processing {ICASSP} 2023, Rhodes Island, Greece, June 4-10, 2023}, pages 1--5. {IEEE}, 2023.

\bibitem[\protect\citeauthoryear{Lee \bgroup \em et al.\egroup }{2024}]{ditspeech}
Keon Lee, Dong~Won Kim, Jaehyeon Kim, and Jaewoong Cho.
\newblock Ditto-tts: Efficient and scalable zero-shot text-to-speech with diffusion transformer.
\newblock {\em CoRR}, abs/2406.11427, 2024.

\bibitem[\protect\citeauthoryear{Leng \bgroup \em et al.\egroup }{2024}]{prompttts2}
Yichong Leng, Zhifang Guo, Kai Shen, Zeqian Ju, Xu~Tan, Eric Liu, Yufei Liu, Dongchao Yang, Leying Zhang, Kaitao Song, Lei He, Xiangyang Li, Sheng Zhao, Tao Qin, and Jiang Bian.
\newblock Prompttts 2: Describing and generating voices with text prompt.
\newblock In {\em The Twelfth International Conference on Learning Representations, {ICLR} 2024, Vienna, Austria, May 7-11, 2024}. OpenReview.net, 2024.

\bibitem[\protect\citeauthoryear{Li \bgroup \em et al.\egroup }{2023}]{bilp2}
Junnan Li, Dongxu Li, Silvio Savarese, and Steven C.~H. Hoi.
\newblock {BLIP-2:} bootstrapping language-image pre-training with frozen image encoders and large language models.
\newblock In Andreas Krause, Emma Brunskill, Kyunghyun Cho, Barbara Engelhardt, Sivan Sabato, and Jonathan Scarlett, editors, {\em International Conference on Machine Learning, {ICML} 2023, 23-29 July 2023, Honolulu, Hawaii, {USA}}, volume 202 of {\em Proceedings of Machine Learning Research}, pages 19730--19742. {PMLR}, 2023.

\bibitem[\protect\citeauthoryear{Lu \bgroup \em et al.\egroup }{2021}]{facevc}
Hsiao{-}Han Lu, Shao{-}En Weng, Ya{-}Fan Yen, Hong{-}Han Shuai, and Wen{-}Huang Cheng.
\newblock Face-based voice conversion: Learning the voice behind a face.
\newblock In Heng~Tao Shen, Yueting Zhuang, John~R. Smith, Yang Yang, Pablo C{\'{e}}sar, Florian Metze, and Balakrishnan Prabhakaran, editors, {\em {MM} '21: {ACM} Multimedia Conference, Virtual Event, China, October 20 - 24, 2021}, pages 496--505. {ACM}, 2021.

\bibitem[\protect\citeauthoryear{Park \bgroup \em et al.\egroup }{2024}]{SYNTHE-SEES}
Jae~Hyun Park, Joon{-}Gyu Maeng, Taejun Bak, and Young{-}Sun Joo.
\newblock {SYNTHE-SEES:} face based text-to-speech for virtual speaker.
\newblock In {\em {IEEE} International Conference on Acoustics, Speech and Signal Processing, {ICASSP} 2024, Seoul, Republic of Korea, April 14-19, 2024}, pages 10321--10325. {IEEE}, 2024.

\bibitem[\protect\citeauthoryear{Radford \bgroup \em et al.\egroup }{2021}]{clip}
Alec Radford, Jong~Wook Kim, Chris Hallacy, Aditya Ramesh, Gabriel Goh, Sandhini Agarwal, Girish Sastry, Amanda Askell, Pamela Mishkin, Jack Clark, Gretchen Krueger, and Ilya Sutskever.
\newblock Learning transferable visual models from natural language supervision.
\newblock In Marina Meila and Tong Zhang, editors, {\em Proceedings of the 38th International Conference on Machine Learning, {ICML} 2021, 18-24 July 2021, Virtual Event}, volume 139 of {\em Proceedings of Machine Learning Research}, pages 8748--8763. {PMLR}, 2021.

\bibitem[\protect\citeauthoryear{Radford \bgroup \em et al.\egroup }{2023}]{whisper}
Alec Radford, Jong~Wook Kim, Tao Xu, Greg Brockman, Christine McLeavey, and Ilya Sutskever.
\newblock Robust speech recognition via large-scale weak supervision.
\newblock In Andreas Krause, Emma Brunskill, Kyunghyun Cho, Barbara Engelhardt, Sivan Sabato, and Jonathan Scarlett, editors, {\em International Conference on Machine Learning, {ICML} 2023, 23-29 July 2023, Honolulu, Hawaii, {USA}}, volume 202 of {\em Proceedings of Machine Learning Research}, pages 28492--28518. {PMLR}, 2023.

\bibitem[\protect\citeauthoryear{Raffel \bgroup \em et al.\egroup }{2020}]{t5}
Colin Raffel, Noam Shazeer, Adam Roberts, Katherine Lee, Sharan Narang, Michael Matena, Yanqi Zhou, Wei Li, and Peter~J. Liu.
\newblock Exploring the limits of transfer learning with a unified text-to-text transformer.
\newblock {\em J. Mach. Learn. Res.}, 21:140:1--140:67, 2020.

\bibitem[\protect\citeauthoryear{Schroff \bgroup \em et al.\egroup }{2015}]{facenet}
Florian Schroff, Dmitry Kalenichenko, and James Philbin.
\newblock Facenet: {A} unified embedding for face recognition and clustering.
\newblock In {\em {IEEE} Conference on Computer Vision and Pattern Recognition, {CVPR} 2015, Boston, MA, USA, June 7-12, 2015}, pages 815--823. {IEEE} Computer Society, 2015.

\bibitem[\protect\citeauthoryear{Sheng \bgroup \em et al.\egroup }{2023}]{Fvmvc}
Zhengyan Sheng, Yang Ai, Yan{-}Nian Chen, and Zhen{-}Hua Ling.
\newblock Face-driven zero-shot voice conversion with memory-based face-voice alignment.
\newblock In Abdulmotaleb El{-}Saddik, Tao Mei, Rita Cucchiara, Marco Bertini, Diana Patricia~Tobon Vallejo, Pradeep~K. Atrey, and M.~Shamim Hossain, editors, {\em Proceedings of the 31st {ACM} International Conference on Multimedia, {MM} 2023, Ottawa, ON, Canada, 29 October 2023- 3 November 2023}, pages 8443--8452. {ACM}, 2023.

\bibitem[\protect\citeauthoryear{Sheng \bgroup \em et al.\egroup }{2024}]{voxeditor}
Zhengyan Sheng, Yang Ai, Li{-}Juan Liu, Jia Pan, and Zhen{-}Hua Ling.
\newblock Voice attribute editing with text prompt.
\newblock {\em CoRR}, abs/2404.08857, 2024.

\bibitem[\protect\citeauthoryear{Shimizu \bgroup \em et al.\egroup }{2024}]{prompttts++}
Reo Shimizu, Ryuichi Yamamoto, Masaya Kawamura, Yuma Shirahata, Hironori Doi, Tatsuya Komatsu, and Kentaro Tachibana.
\newblock Prompttts++: Controlling speaker identity in prompt-based text-to-speech using natural language descriptions.
\newblock In {\em {IEEE} International Conference on Acoustics, Speech and Signal Processing, {ICASSP} 2024, Seoul, Republic of Korea, April 14-19, 2024}, pages 12672--12676. {IEEE}, 2024.

\bibitem[\protect\citeauthoryear{Tan \bgroup \em et al.\egroup }{2022}]{ns}
Xu~Tan, Jiawei Chen, Haohe Liu, Jian Cong, Chen Zhang, Yanqing Liu, Xi~Wang, Yichong Leng, Yuanhao Yi, Lei He, Frank~K. Soong, Tao Qin, Sheng Zhao, and Tie{-}Yan Liu.
\newblock Naturalspeech: End-to-end text to speech synthesis with human-level quality.
\newblock {\em CoRR}, abs/2205.04421, 2022.

\bibitem[\protect\citeauthoryear{Tong \bgroup \em et al.\egroup }{2023}]{ot-cfm}
Alexander Tong, Nikolay Malkin, Guillaume Huguet, Yanlei Zhang, Jarrid Rector{-}Brooks, Kilian Fatras, Guy Wolf, and Yoshua Bengio.
\newblock Conditional flow matching: Simulation-free dynamic optimal transport.
\newblock {\em CoRR}, abs/2302.00482, 2023.

\bibitem[\protect\citeauthoryear{Vyas \bgroup \em et al.\egroup }{2023}]{audiobox}
Apoorv Vyas, Bowen Shi, Matthew Le, Andros Tjandra, Yi{-}Chiao Wu, Baishan Guo, Jiemin Zhang, Xinyue Zhang, Robert Adkins, William Ngan, Jeff Wang, Ivan Cruz, Bapi Akula, Akinniyi Akinyemi, Brian Ellis, Rashel Moritz, Yael Yungster, Alice Rakotoarison, Liang Tan, Chris Summers, Carleigh Wood, Joshua Lane, Mary Williamson, and Wei{-}Ning Hsu.
\newblock Audiobox: Unified audio generation with natural language prompts.
\newblock {\em CoRR}, abs/2312.15821, 2023.

\bibitem[\protect\citeauthoryear{Wang \bgroup \em et al.\egroup }{2023a}]{valle}
Chengyi Wang, Sanyuan Chen, Yu~Wu, Ziqiang Zhang, Long Zhou, Shujie Liu, Zhuo Chen, Yanqing Liu, Huaming Wang, Jinyu Li, Lei He, Sheng Zhao, and Furu Wei.
\newblock Neural codec language models are zero-shot text to speech synthesizers.
\newblock {\em CoRR}, abs/2301.02111, 2023.

\bibitem[\protect\citeauthoryear{Wang \bgroup \em et al.\egroup }{2023b}]{cam++}
Hui Wang, Siqi Zheng, Yafeng Chen, Luyao Cheng, and Qian Chen.
\newblock {CAM++:} {A} fast and efficient network for speaker verification using context-aware masking.
\newblock In Naomi Harte, Julie Carson{-}Berndsen, and Gareth Jones, editors, {\em 24th Annual Conference of the International Speech Communication Association, Interspeech 2023, Dublin, Ireland, August 20-24, 2023}, pages 5301--5305. {ISCA}, 2023.

\bibitem[\protect\citeauthoryear{Wang \bgroup \em et al.\egroup }{2024}]{atr}
Qian Wang, Jia{-}Chen Gu, and Zhen{-}Hua Ling.
\newblock Multiscale matching driven by cross-modal similarity consistency for audio-text retrieval.
\newblock In {\em {IEEE} International Conference on Acoustics, Speech and Signal Processing, {ICASSP} 2024, Seoul, Republic of Korea, April 14-19, 2024}, pages 11581--11585. {IEEE}, 2024.

\bibitem[\protect\citeauthoryear{Weng \bgroup \em et al.\egroup }{2023}]{SP-FaceVC}
Shao{-}En Weng, Hong{-}Han Shuai, and Wen{-}Huang Cheng.
\newblock Zero-shot face-based voice conversion: Bottleneck-free speech disentanglement in the real-world scenario.
\newblock In Brian Williams, Yiling Chen, and Jennifer Neville, editors, {\em Thirty-Seventh {AAAI} Conference on Artificial Intelligence, {AAAI} 2023, Thirty-Fifth Conference on Innovative Applications of Artificial Intelligence, {IAAI} 2023, Thirteenth Symposium on Educational Advances in Artificial Intelligence, {EAAI} 2023, Washington, DC, USA, February 7-14, 2023}, pages 13718--13726. {AAAI} Press, 2023.

\bibitem[\protect\citeauthoryear{Yang \bgroup \em et al.\egroup }{2024}]{instructtts}
Dongchao Yang, Songxiang Liu, Rongjie Huang, Chao Weng, and Helen Meng.
\newblock Instructtts: Modelling expressive {TTS} in discrete latent space with natural language style prompt.
\newblock {\em {IEEE} {ACM} Trans. Audio Speech Lang. Process.}, 32:2913--2925, 2024.

\bibitem[\protect\citeauthoryear{Yao \bgroup \em et al.\egroup }{2024}]{promptvc}
Jixun Yao, Yuguang Yang, Yi~Lei, Ziqian Ning, Yanni Hu, Yu~Pan, Jingjing Yin, Hongbin Zhou, Heng Lu, and Lei Xie.
\newblock Promptvc: Flexible stylistic voice conversion in latent space driven by natural language prompts.
\newblock In {\em {IEEE} International Conference on Acoustics, Speech and Signal Processing, {ICASSP} 2024, Seoul, Republic of Korea, April 14-19, 2024}, pages 10571--10575. {IEEE}, 2024.

\bibitem[\protect\citeauthoryear{Ye \bgroup \em et al.\egroup }{2024}]{asq}
Lingxuan Ye, Changfeng Gao, Gaofeng Cheng, Liuping Luo, and Qingwei Zhao.
\newblock {ASQ:} an ultra-low bit rate asr-oriented speech quantization method.
\newblock {\em {IEEE} Signal Process. Lett.}, 31:221--225, 2024.

\bibitem[\protect\citeauthoryear{Zhang \bgroup \em et al.\egroup }{2023}]{promptspeaker}
Yongmao Zhang, Guanghou Liu, Yi~Lei, Yunlin Chen, Hao Yin, Lei Xie, and Zhifei Li.
\newblock Promptspeaker: Speaker generation based on text descriptions.
\newblock In {\em {IEEE} Automatic Speech Recognition and Understanding Workshop, {ASRU} 2023, Taipei, Taiwan, December 16-20, 2023}, pages 1--7. {IEEE}, 2023.

\end{thebibliography}

\end{document}